\documentclass[reprint,showpacs,amsmath,amssymb,aps]{revtex4-1}
\usepackage{mathrsfs}
\usepackage{amsfonts,gensymb}
\usepackage{amsmath,amsfonts,amssymb,mathrsfs,bm}
\usepackage{verbatim,psfrag,times,CJK}
\usepackage{graphicx,graphics,color,epsfig}
\usepackage{float}
\usepackage{flafter}
\usepackage{subeqnarray}
\usepackage{cases}
\usepackage{amsmath, upgreek, amssymb, graphicx, subfigure, paralist, dsfont}
\usepackage[colorlinks, linkcolor=blue, citecolor=blue, urlcolor=blue, breaklinks=true]{hyperref}

\definecolor{red}{rgb}{1,0,0}
\newcommand{\ket}[1]{\left\vert{#1}\right\rangle}
\newcommand{\bra}[1]{\left\langle{#1}\right\vert}

\begin{document}

\title{Highly directional photon superbunching from a `few-atom'  chain of emitters}

\author{Qurrat-ul-Ain \surname{Gulfam}}
\email{qgulfam@jazanu.edu.sa} 
\affiliation{Department of Physics, Faculty of Science, Jazan University, P.O. Box 114, Gizan 45142, Saudi Arabia}
\author{Zbigniew \surname{Ficek}}
\email{zficek@kacst.edu.sa} 
\affiliation{The National Centre for Applied Physics, KACST, P.O. Box 6086, Riyadh 11442, Saudi Arabia}
\affiliation{Quantum Optics and Engineering Division, Institute of Physics, University of Zielona G\'ora, Szafrana 4a, Zielona G\'ora 65-516, Poland}

\date{\today}

\begin{abstract}
We examine angular distribution of the probability of correlated fluorescence photon emission from a linear chain of identical equidistant two-level atoms. We selectively excite one of the atoms by a resonant laser field. The atoms are coupled to each other via the dipole-dipole interaction and collective spontaneous emission. Our attention is focused on the simultaneous observation of correlated pairs of photons. It is found that the interference between the emitting atoms can result in a highly directional emission of photon pairs. These pairs of photons posses strong correlations and their emission is highly concentrated into specific detection directions. We demonstrate the crucial role of the selective coherent excitation in such a geometrical configuration. Shifting the driving field from an atom located at one end of the chain to the other causes the radiation pattern to flip to the opposite half of the detection plane. Furthermore, we find that atomic systems in which only an atom situated at a particular position within the linear chain is driven by a laser field can radiate correlated twin photons in directions along which the radiation of single photons is significantly reduced.  Alternatively, superbunching in the emitted photon statistics preferentially occurs in directions of negligible or vanishing single photon emission. The effect of superbunching strengthens as more emitters are added to the chain. Depending on the number of atoms and the position of the driven atom within the chain, the strongly correlated pairs of photons can be emitted into well-defined single, two or four directions.
 
\end{abstract}

\pacs{37.10.Jk, 42.25.Fx, 42.25.Hz, 42.50.Gy}

\maketitle

\section{Introduction}

Correlated systems is an area of active research in quantum optics. Ultracold atoms prove to be a potential candidate for the development of few-atom atomic chains~\cite{coldgas,Ni,ryd1,ryd2,nm07,vr10,coldgasImaging}.  Correlation functions of the electromagnetic field radiated by atomic systems are often adopted to determine nonclassical and entangled properties of quantum optical systems~\cite{agarwal,ficekbook}. The first-order correlation functions are used to determine the intensity and spectral properties of the radiated field~\cite{scullybook}. Higher order correlation functions, in particular, the second-order correlation function is used to determine whether the radiated field is quantum or classical in nature, or in other words, if two simultaneously emitted photons are correlated or anticorrelated.

From its first-ever development~\cite{d54} to present, the collective behavior of spatially separated atoms coupled to each other through the vacuum-induced dipole dipole interaction has been extensively studied~\cite{Lehmberg,bb84,f86}. Numerous papers discuss this collective radiation pattern dependent on a specific spatial geometry~\cite{corr1,corr2,corr3,corr4,distance,Carmichael, mf07,Cirac,lz14}. The collective behavior of dipole dipole interacting atoms can also be examined in terms of the correlation functions.

Recently, we have shown that the collective behavior of a linear chain of atoms can result in a strong directional emission of photons into well-defined modes~\cite{mirror}. The strong directivity has been predicted without the requirement of the coupling of the system to some external medium, for example, a nano-wave guide or a fibre. 

High directivity can be observed in the higher order correlations. For example, Richter analyzed the dependence on the pumping process of the second-order correlation function of the field emitted from three atoms and found that the directionality of the correlations is different for coherent and incoherent pumping~\cite{Richter, richter2}. 

The normalized second-order correlation function $g^{(2)}$ can have different values. Photon statistics are determined from these values. Values of $g^{(2)}<1$  refer to as photon anti-bunching, $1<g^{(2)}<2$ as bunching, and $g^{(2)}>2 $ as superbunching. Photon superbunching, the simultaneous detection of multi photons and antibunching, the emission of single photons in a regular manner are  subjects of widespread attention~\cite{walls, kl76,superbunching1, characterization1, superbunching2,kl77,antibunching1,antibunching2,antibunching3,antibunching4,antibunching5,antibunching6,antibunching2,Wiegand,Wstate,Bin,Leuchs,AIP-paper}. Recently, photon antibunching has been tested at the level of a single atom placed in a cavity~\cite{antibunching1,antibunching2,antibunching3,antibunching4,antibunching5,antibunching6}. Superbunching in resonance fluorescence from two atoms is discussed in~\cite{Wiegand}. It is set forth as a tool to study non-classical effects such as atom-field entanglement~\cite{superbunching1,characterization1, superbunching2,AIP-paper} . Both effects for non-interacting two-level atoms have been inspected~\cite{Wstate}. Interferometric studies of photon superbunching of classical light have been carried out in~\cite{Bin}. Superbunching for a squeezed vacuum state is studied in~\cite{Leuchs}.

In this paper, we study the one-time second-order correlation function of fluorescence photons emitted by a linear chain of identical two-level atoms and detected by a single photo detector in the far field zone. We assume that only one of the atoms is driven on resonance with an external laser field. The atoms are damped by their collective coupling to the common vacuum electromagnetic field. The angular distribution of the correlation functions is studied for equal spacing between the atoms.
By monitoring the angular distribution of the correlation functions of the emitted photons we can determine the correlation properties of a particular pair of atoms of a linear atomic chain.
Our primary findings are that for near by placed atoms, the selective excitation of one of the atoms situated at either end or mid-way the atomic chain plays a pivotal role in determining directions of emission of correlated photons. Single or two specific detection directions (depending on the number of atoms) along which highly correlated photons are emitted have been observed when an atom fixed at one end of the chain is driven. These directions are always found to be located in the opposite half of the observation plane relative to the half where the driven atom is present. Interestingly, exciting the middle atom of the chain by the laser field produces two or four such prominent directions. This is because the system divides itself into two constituent sub-chains; the driven atom is present at one end for one half sub-chain, and at the opposite end for the other half sub-chain. 
We also show that the effect of superbunching can be enhanced by increasing the number of atoms, $N$ in the chain. This enhancement is not only limited to the increase in the magnitude of the probability of emission of twin photons but also the detection directions become more and more precise. Intriguing is the fact that one has to distinguish here if $N$ is an even or odd number. Adding more atoms to the chain while keeping the inter atom distance constant diverts the directions of emission of strongly correlated photons toward the atomic axis.
Last but not the least, we reveal that first order correlation function, a measure of intensity, holds an influential place in determining the normalized second order correlation function. In other words, the normalized intensity-intensity correlation function has been noticed to contain more information about the emission of single photons than about the correlated two-photon emission.
\begin{figure}[h]
  \centerline{\includegraphics[width=\columnwidth]{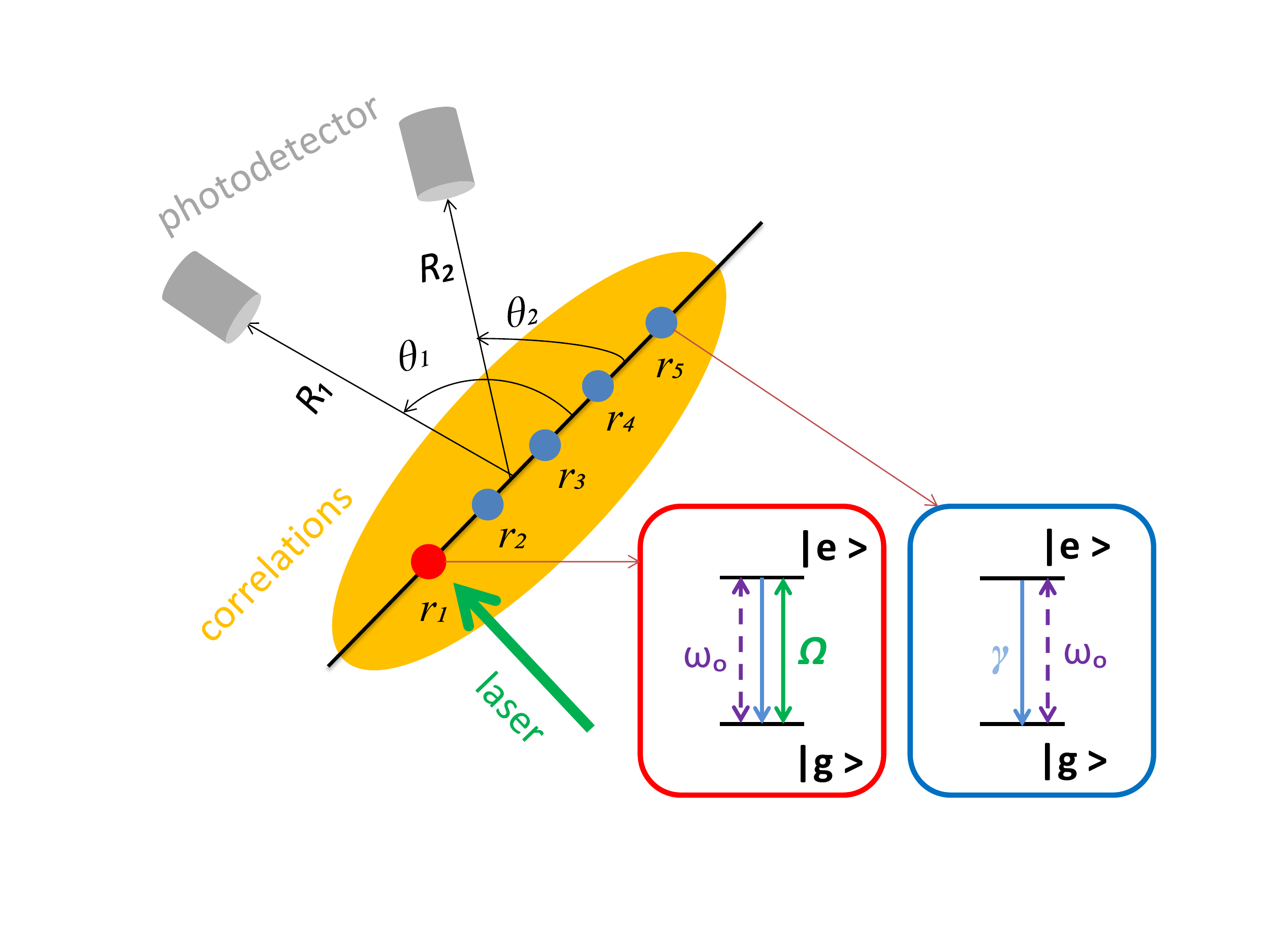}}
  \caption{A chain of identical equidistant  atoms. An external laser field of Rabi frequency $\Omega$ drives one of the atoms. The atoms are coupled to each other through the dipole-dipole interaction and the collective spontaneous emission resulting from the coupling of the atoms to a common vacuum field. The emitted fluorescence field is detected by 
a single or two photo detectors at distances $R_1$ and $R_2$ and angles $\theta_1$ and $\theta_2$ from the atomic line, located in the plane normal to the plane defined by the polarization of the atomic dipole moments $\vec{\mu}\cdot \hat{r}_{ij}=0$.  Red (blue) blob and inset shows a coherently driven (non-driven) atom.}
  \label{fig1}
\end{figure}

The paper is organized as follows. Sec.~\ref{sec1} explains our model system. Two-level atoms are arranged along a line in the form of a chain. A resonant laser drives one of the atoms. Photo detection aims at spotting correlated photon pairs. In Sec.~\ref{sec2}, we give definitions of the first- and second-order correlation functions in terms of the atomic dipole operators. The master equation approach for the evaluation of  density matrix elements of the atomic system is outlined. The stationary state solution of the master equation allows to calculate the intensity of the scattered light and the intensity-intensity correlation function in the steady state realm. Sec.~\ref{sec3} discusses the angular distribution of the correlation functions for different number of atoms in the chain in detail. Parameter regimes and values of detection angles have been identified for the observation of highly correlated photon doublets. The critical effect of switching the coherent driving field for atoms positioned at the right or left or at the middle of the chain is illustrated for different number of atoms. We augment our analytical analysis by presenting polar plots for the probability patterns of two photon emission and comparing this correlated two photon emission diagrams to that of single uncorrelated photons. The effect of increasing the number of atoms in the chain on the magnitude and important detection directions for the observation of superbunched photons have been highlighted. Sec.~\ref{sec4} describes the physical meaning of superbunching for correlated florescence light quanta. Another more meaningful measurement for two-photon correlations is adopted. Finally, in Sec.~\ref{sec5} a succinct account of the major findings of the paper is given.

\section{Model system}
\label{sec1}

The system under investigation consists of a set of $N$ identical two-level atoms. The atoms are equally spaced and are placed at fixed positions $r_i$ along a line such that they form a linear chain, as shown in Fig.~\ref{fig1}. They are supposed to be in a close spatial proximity to each other such that they couple to a common three-dimensional vacuum field and interact via the so called vacuum-induced dipole-dipole interaction mechanism. An incident continuous wave laser field (shown in green) with frequency equal to the atomic transition frequency, $\omega_{0}$ drives one of the atoms in the chain. In particular, the left-most, or the middle or the right-most atom is driven at a time. $\Omega$ is the Rabi frequency associated with the laser field and its propagation vector lies perpendicular to the atomic line.  Our aim is to record highly correlated simultaneously emitted photon pairs from the light scattered off the atomic ensemble. For this purpose, two photo detectors located at distances $R_1$ and $R_2$, and at polar angles $\theta_1$ and $\theta_2$, respectively,  in the far-field region detect the scattered light. Photo detection takes place in a plane orthogonal to the direction of alignment of atomic dipole moments. The left (red) inset shows the internal energy-level structure corresponding to a laser-driven atom (red blob) in which the atomic transition is coupled with the laser field while right (blue) inset refers to any atom which is not being driven (blue blob).  $\ket e$ ($\ket g$) represents the excited (ground) state of a bi-level atom. The energy separation between the excited and the ground states is given by $\hbar\omega_0$.

\section{Correlation functions}\label{sec2}

We focus on correlation properties of the photons emitted by an atomic system, which in turn, can be used to determine the correlation properties of the atoms. In order to do this, we introduce first-order correlation function of the normally ordered electric field operators associated with the fluorescence field emitted by the atoms and detected in the far-field zone~\cite{glauber,scullybook}
\begin{align}
G^{(1)}(\vec{R},t) &= \left(\frac{R^{2}}{2\pi k_{0}}\right)\left\langle \vec{E}^{(-)}(\vec{R},t)\cdot\vec{E}^{(+)}(\vec{R},t)\right\rangle ,\label{q1}
\end{align}
and the normally ordered intensity correlation function
\begin{align}
&G^{(2)}(\vec{R_{1}},t_{1};\vec{R}_{2},t_{2}) = \langle: I(\vec{R}_{1},t_{1}) I(\vec{R}_{2},t_{2}):\rangle \nonumber\\
&=\left(\frac{R_{1}R_{2}}{2\pi k_{0}}\right)^{2}\langle \vec{E}^{(-)}(\vec{R}_{1},t_{1})\vec{E}^{(-)}(\vec{R}_{2},t_{2}) \nonumber\\
&\times \vec{E}^{(+)}(\vec{R}_{2},t_{2})\vec{E}^{(+)}(\vec{R}_{1},t_{1})\rangle ,\label{q2}
\end{align}
where $\vec{E}^{(+)} (\vec{E}^{(-)})$ denotes the positive (negative) frequency part of the electric field. In the definition of the correlation function $G^{(1)}(\vec{R},t)$, Eq.~(\ref{q1}),  we have introduced the factor $(R^{2}/2\pi k_{0})$ so that $G^{(1)}(\vec{R},t)d\Omega_{R}dt$ is the probability of finding a photon inside the solid angle $d\Omega_{R}$ around the direction $\vec{R}$ in the time interval $dt$. 

Assume that the system of radiating atoms is composed of $N$ equidistant identical two-level atoms. An atom, say the $i$th one, is represented by its ground state $\ket{g_{i}}$, an excited state $\ket{e_{i}}$, the Bohr atomic transition frequency $\omega_{0}$, the transition dipole moment $\vec{\mu}$, and its position $\vec{r}_{i}$ along the atomic line.
In the far-field zone of the radiating atoms, the contribution from the free field can be neglected, and the positive frequency part of the scattered electric field can be expressed in terms of the transition dipole moments of the atoms as
\begin{align}
\vec{E}^{(+)}(\vec{R},t) &= \frac{\omega_{0}^{2}\mu^{2}}{4\pi\epsilon_{0}c^{2}}
\frac{[(\hat{R}\times\hat{\mu})\times\hat{R}]}{R}\sum_{i=1}^{N} S_{i}^{-}{\rm e}^{-i(k\hat{R}\cdot \vec{r}_{i}-\omega_{0}t)} ,\label{q3}
\end{align}
where $S_{i}^{-}=\ket{g_{i}}\bra{e_{i}}$ is the atomic lowering operator for the $i$th atom, $\mu =|\bra{g_{i}}\vec{\mu}\ket{e_{i}}|$ is the magnitude of the atomic dipole moment, $R=|\vec{R}|$, $\hat{\mu}$ is the unit vector in the direction of the atomic transition dipole moment, and $\hat{R}$ is the unit vector in the direction of observation $\vec{R}$. 

After substituting Eq.~(\ref{q3}) into Eqs.~(\ref{q1}) and (\ref{q2}), we obtain
\begin{align}
G^{(1)}(\vec{R},t) &= u(\hat{R})\,\gamma\sum_{\{i,j\}=1}^{N}\langle S_{i}^{+}(t)S_{j}^{-}(t)\rangle e^{ik\vec{r}_{ij}\cdot\hat{R}} ,\label{q4}
\end{align}
and
\begin{align}
&G^{(2)}(\vec{R_{1}},t_{1};\vec{R}_{2},t_{2}) = u(\hat{R}_{1})u(\hat{R}_{2})\,\gamma^{2} \nonumber\\
&\sum_{i\neq j=1}^{N}\sum_{l\neq k=1}^{N}\langle S_{i}^{+}(t_{1})S_{j}^{+}(t_{2})S_{k}^{-}(t_{2})S_{l}^{-}(t_{1})\rangle \nonumber\\
&\times \exp\left[ik\left(\vec{r}_{il}\cdot\hat{R}_{1}+\vec{r}_{jk}\cdot\hat{R}_{2}\right)\right] ,\label{q5}
\end{align}
where $\vec{r}_{ij}=\vec{r}_{i}-\vec{r}_{j}$ is the spatial distance vector between the atoms $i$ and $j$, $u(\hat{R})=(3/8\pi)[1-(\hat{\mu}\cdot\hat{R})^{2}]$ is the radiation pattern of a single atomic dipole, and $\gamma$ is the decay rate of the atomic transition.

The evolution of the atomic correlation functions appearing in Eqs.~(\ref{q4}) and (\ref{q5}) is governed by a master equation for the atomic density operator $\rho$. Within the Born-Markov and the rotating wave approximation, the density operator satisfies the Lehmberg-Agarwal master equation~\cite{fs,agarwal}
\begin{align}
\frac{\partial \rho}{\partial t} =& -\frac{i}{\hbar}\left[H,\rho\right]  -i\sum_{i\neq j=1}^{N}\Omega_{ij}\!\left[S^{+}_{i}S^{-}_{j},\rho\right]\nonumber\\
&- \frac{1}{2}\sum_{i,j=1}^{N}\gamma_{ij}\left(S_{i}^{+}S_{j}^{-}\rho+\rho S_{i}^{+}S_{j}^{-} -2S_{j}^{-}\rho S_{i}^{+}\right) ,\label{q6}
\end{align}
where
\begin{align}
H &= \hbar\,\omega_{0}\sum_{i=1}^{N}S^{+}_{i}S^{-}_{i} +\frac{1}{2}\hbar\Omega\left(S_{l}^{+}e^{-i\omega_{0}t} +S_{l}^{-}e^{i\omega_{0}t}\right) ,\label{q7}
\end{align}
\begin{align}
\gamma_{ij} &= \frac{3}{2}\gamma\left\{ \left[1-\left( \hat{\mu}\cdot \hat{r}_{ij}\right)^{2} \right]
 \frac{\sin\xi_{ij}}{\xi_{ij}}\right.  \nonumber \\
&\left. +\left[ 1-3\left( \hat{\mu}\cdot\hat{r}_{ij}\right)^{2} \right]\left[ \frac{\cos\xi_{ij} }
{\xi_{ij}^{2}}-\frac{\sin\xi_{ij} }{\xi_{ij}^{3}}\right] \right\} ,\label{q8}\\
\Omega_{ij} &= \frac{3}{4}\gamma\left\{ -\left[1-\left( \hat{\mu}\cdot \hat{r}_{ij}\right)^{2} \right] 
\frac{\cos\xi_{ij}}{\xi_{ij}}\right.  \nonumber \\
&\left. +\left[ 1-3\left(\hat{\mu}\cdot\hat{{r}}_{ij}\right)^{2} \right]\left[\frac{\sin\xi_{ij} }
{\xi_{ij}^{2}} +\frac{\cos\xi_{ij}}{\xi_{ij}^{3}}\right]\right\} ,\label{q9}
\end{align}
in which $\gamma\equiv\gamma_{ii}$ is the spontaneous emission rate for each individual atom, and
\begin{align}
\xi_{ij} &=\frac{2\pi r_{ij}}{\lambda} ,\quad r_{ij}\equiv |\vec{r}_{ij}|= |\vec{r}_{j}-\vec{r}_{i}| .\label{q10}
\end{align}
The $\gamma_{ij}$ terms describe the collective damping which results from an incoherent exchange of photons between the atoms $i$ and $j$, and the $\Omega_{ij}$ terms describe the collective shift of the atomic energy levels. The shift results from a coherent exchange of photons, the dipole-dipole interaction between the atoms. The effect of $\Omega_{ij}$ on the atomic system is the shift of the energy of the collective states from the single-atom energy states. 
We have chosen the laser field with Rabi frequency $\Omega$ to drive only one, the $l$-th atom of the chain, and to be exactly resonant with the atomic transition frequency. With this driving arrangement, an excitation can be transferred between the atoms through the dipole-dipole interaction and collective damping of the atoms.

We solve the master equation, Eq.~(\ref{q6})  numerically up to $N=3$ atoms and calculate the steady-state values of the correlation functions. We investigate angular distributions of the correlation functions and analyze their dependence on the way an atom at a particular position in the chain  is excited.

\section{Angular distributions of the correlations}\label{sec3} 

Let us now demonstrate how one could determine the correlation properties between two atoms of a chain composed of $N\geq 2$ atoms by monitoring the angular distribution of the correlation functions of the emitted photons. We illustrate this in detail with examples of atomic chains composed of two and three atoms and then consider the case of $N>3$. We determine the general conditions for the angular distribution of the correlation functions and its dependence on the number of atoms and detectors.

\subsection{Atomic chain composed of two atoms}

Consider first the simplest case, a chain composed of only two atoms and calculate the first-order correlation function of the steady-state radiation field, $G^{(1)}(\vec{R})\equiv \lim_{t\rightarrow\infty}G^{(1)}(\vec{R},t)$. The correlation function gives us the information about the probability of emitting single photons in the direction of detection $\vec{R}$. It also describes the intensity of the radiation field emitted in the observation direction $\vec{R}$. The general properties of the first order correlation function can be determined from Eq.~(\ref{q4}), which can be written as
\begin{align}
G^{(1)}(\vec{R}) &= u(\hat{R})\,\gamma\,\!\left\{\langle S_{1}^{+}S_{1}^{-}\rangle +\langle S_{2}^{+}S_{2}^{-}\rangle\right. \nonumber\\
&\left. +2\,{\rm Re}[\langle S_{1}^{+}S_{2}^{-}\rangle]\cos\left(k\,r_{12}\cos\theta\right)\right. \nonumber\\
&\left. +2\,{\rm Im}[\langle S_{1}^{+}S_{2}^{-}\rangle]\sin\left(k\,r_{12}\cos\theta\right)\right\} ,\label{q11}
\end{align}
where $\theta$ is the angle between the interatomic axis and the direction of observation.
The expression (\ref{q11}) can be written in a compact form as
\begin{align}
G^{(1)}(\vec{R}) &= u(\hat{R})\gamma\left(I_{1}+I_{2}\right) \nonumber\\
&\times\left[1+\upsilon_{12}\cos\left(k\,r_{12}\cos\theta -\psi_{12}\right)\right] ,\label{q12}
\end{align}
where $I_{i}=\langle S_{i}^{+}S_{i}^{-}\rangle$ is the intensity of light emitted by atom~$i$,  
\begin{align}
\upsilon_{12} = \frac{2|\langle S_{1}^{+}S_{2}^{-}\rangle|}{\langle S_{1}^{+}S_{1}^{-}\rangle +\langle S_{2}^{+}S_{2}^{-}\rangle}  \label{q13}
\end{align}
is the first-order coherence between the atoms 1 and 2, and $\psi_{12}={\rm arg}(\langle S_{1}^{+}S_{2}^{-}\rangle)$. The angle $\psi_{12}$ depends on the sign of the real and imaginary parts of $\langle S_{1}^{+}S_{2}^{-}\rangle$ such that
\begin{align}
\psi_{12} &= \tan^{-1}\left(\frac{{\rm Im}\langle S_{1}^{+}S_{2}^{-}\rangle}{{\rm Re}\langle S_{1}^{+}S_{2}^{-}\rangle}\right) ,
\end{align}
when ${\rm Re}\langle S_{1}^{+}S_{2}^{-}\rangle>0, {\rm Im}\langle S_{1}^{+}S_{2}^{-}\rangle > 0$.
\begin{align}
\psi_{12} &= \pi -\tan^{-1}\left(\frac{{\rm Im}\langle S_{1}^{+}S_{2}^{-}\rangle}{{\rm Re}\langle S_{1}^{+}S_{2}^{-}\rangle}\right) ,
\end{align}
when ${\rm Re}\langle S_{1}^{+}S_{2}^{-}\rangle < 0, {\rm Im}\langle S_{1}^{+}S_{2}^{-}\rangle > 0$.
\begin{align}
\psi_{12} &= -\tan^{-1}\left(\frac{{\rm Im}\langle S_{1}^{+}S_{2}^{-}\rangle}{{\rm Re}\langle S_{1}^{+}S_{2}^{-}\rangle}\right) ,
\end{align}
when ${\rm Re}\langle S_{1}^{+}S_{2}^{-}\rangle > 0, {\rm Im}\langle S_{1}^{+}S_{2}^{-}\rangle < 0$, and
\begin{align}
\psi_{12} &= -\pi +\tan^{-1}\left(\frac{{\rm Im}\langle S_{1}^{+}S_{2}^{-}\rangle}{{\rm Re}\langle S_{1}^{+}S_{2}^{-}\rangle}\right) ,
\end{align}
when ${\rm Re}\langle S_{1}^{+}S_{2}^{-}\rangle < 0, {\rm Im}\langle S_{1}^{+}S_{2}^{-}\rangle < 0$.

Viewing as a function of the detection angle $\theta$, we see that in general the angular distribution of $G^{(1)}(\vec{R})$ is not spherically symmetric. The correlation function exhibits an interference pattern dependent not only on the separation between the atoms but also on the populations of the atoms and the coherences between them. In other words, the radiation intensity pattern depends on the way the atoms are excited. When the atoms are prepared in a state or driven such that the correlation function $\langle S_{1}^{+}S_{2}^{-}\rangle$ is real the modulation of the angular distribution depends solely on $r_{12}$. However, when the correlation function is complex, then not only the distance between the atoms but also the inter atomic correlation function plays a crucial role in the angular distribution of the emitted photons. Thus, the depth of the modulation, determined by the cosine term, not only depends on the geometry of the system, determined by $r_{12}$, but also on the way the atoms are correlated, which is determined by $\upsilon_{12}$ and $\psi_{12}$.

Let us find the values of the detection angle $\theta$ at which $G^{(1)}(\vec{R})$ can be maximal or minimal, corresponding to directions of a strong focusing or divergence of the emitted photons, and how these observation angles depend on $r_{12}$ and $\psi_{12}$. Note that directions in which $G^{(1)}(\vec{R})$ is zero or close to zero correspond to elimination of  single photon emission in those directions. To find these observation angles we evaluate $\partial G^{(1)}/\partial\theta$ and get
\begin{align}
\frac{\partial G^{(1)}}{\partial\theta} &= u(\hat{R})\gamma\left(I_{1}+I_{2}\right)\upsilon_{12}\,k\,r_{12} \nonumber\\
&\times \sin\left(k\,r_{12}\cos\theta -\psi_{12}\right)\sin\theta .
\end{align}
Next,
\begin{align}
\frac{\partial G^{(1)}}{\partial\theta} =0 \Leftrightarrow   \sin\theta =0 \ \ {\rm or}\ \ \sin\left(k\,r_{12}\cos\theta -\psi_{12}\right)=0 .
\end{align}
Hence, the values of $\theta$ at which  $G^{(1)}(\vec{R})$ is maximal or minimal are
\begin{align}
\theta = n\pi \quad {\rm or}\quad \theta = \arccos\left(\frac{n\pi +\psi_{12}}{k\,r_{12}}\right) , \quad n\in\{0, \pm 1,\pm 2\}. \label{q14}
\end{align}
From this it follows that there are two separate criteria for $\theta$ at which maxima and minima of $G^{(1)}(\vec{R})$ could occur. The criterion $\theta=n\pi$ is independent of the distance between the atoms and the angle $\psi_{12}$. Therefore, there always can be either maximum or minimum of $G^{(1)}(\vec{R})$ along the atomic axis. The second criterion shows that there can be maxima and minima where the angular distribution depends on both, the distance between the atoms and $\psi_{12}$. Clearly, if $\langle S_{1}^{+}S_{2}^{-}\rangle$ has zero imaginary part then the angular distribution of $G^{(1)}(\vec{R})$ depends solely on $r_{12}$. Thus, the interference is not purely geometrical it also depends on the manner the atoms are correlated. 

From Eq.~(\ref{q14}) it is also seen that maxima and minima may appear in directions other than the direction of the atomic axis if
\begin{align}
r_{12}/\lambda \geq \frac{1}{2}|(n+\psi_{12}/\pi)| .
\end{align}
Note that for the pure geometrical case of $\psi_{12}=0$ and atomic separations $r_{12}<\lambda/2$, the condition for the optimum negative value (that is $-1$) of the term $\cos(k\,r_{12}\cos\theta)$ necessary to achieve optimal reduction of $G^{(1)}(\vec{R})$ cannot be achieved for all values of $\theta$. In physical terms, at distances $r_{12}<\lambda/2$ the atomic dipole moments oscillate in phase resulting in an enhanced emission of photons. 
However, the term $\cos(kr_{12}\cos\theta -\psi_{12})$ can reach the optimum negative value for $r_{12}<\lambda/2$, i.e., a nonzero $\psi_{12}$ can lift the limit. 
For example, in the case of $\psi_{12}=-3\pi/4$ corresponding to ${\rm Re}\langle S_{1}^{+}S_{2}^{-}\rangle={\rm Im}\langle S_{1}^{+}S_{2}^{-}\rangle<0$, and $r_{12}=\lambda/4$, there are two directions at which $\cos(kr_{12}\cos\theta -\psi_{12})=-1$, namely $\theta=\pi/6$ and $\theta=5\pi/3$. At these two directions $G^{(1)}(\vec{R})$ can be optimally reduced. It is interesting that a change of the sign of ${\rm Im}\langle S_{1}^{+}S_{2}^{-}\rangle$ from negative to positive results in a rotation of those two directions by $\pi$. In physical terms, a nonzero $\psi_{12}$ can shift the phase difference between the atomic dipole moments such that the dipoles could oscillate in an opposite phase resulting in a reduction or even inhibition of the emission of photons. The inhibition of the single photon emission may occur only when $\upsilon_{12}=1$, i.e., when the oscillations of the atomic dipole moments are perfectly coherent. 
\begin{figure}[ht]
  \includegraphics[width=.48\columnwidth]{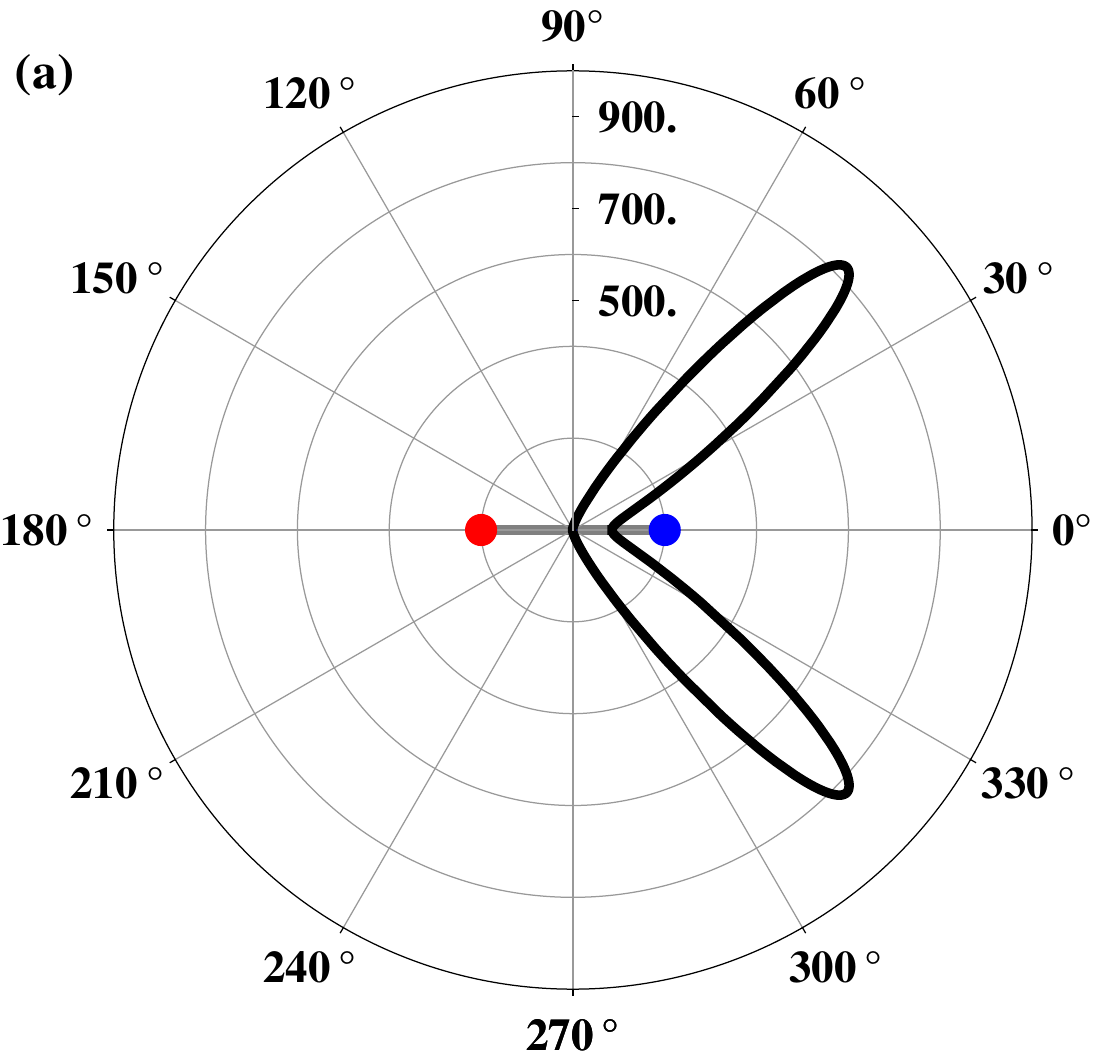}
 \includegraphics[width=.48\columnwidth]{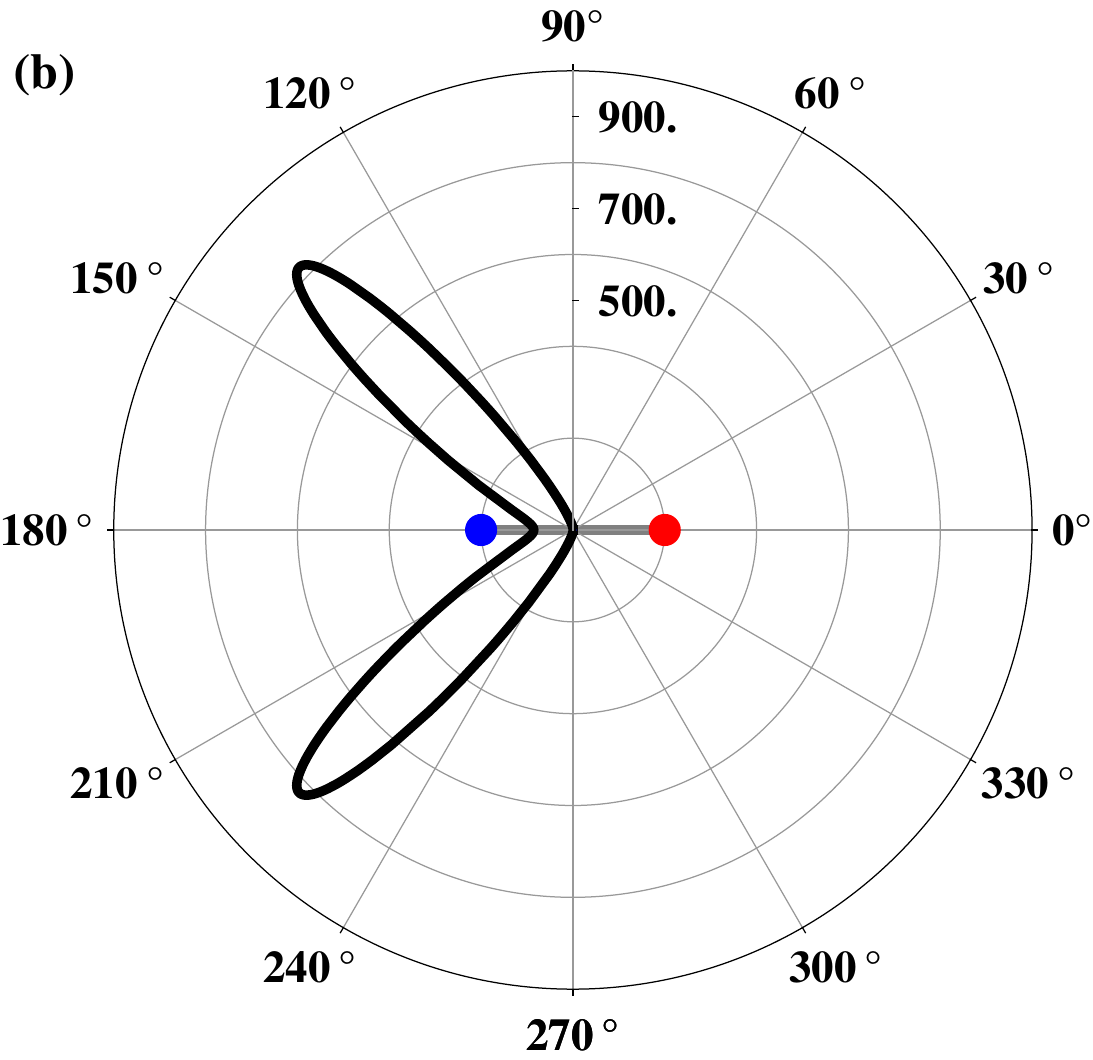}
 \includegraphics[width=.48\columnwidth]{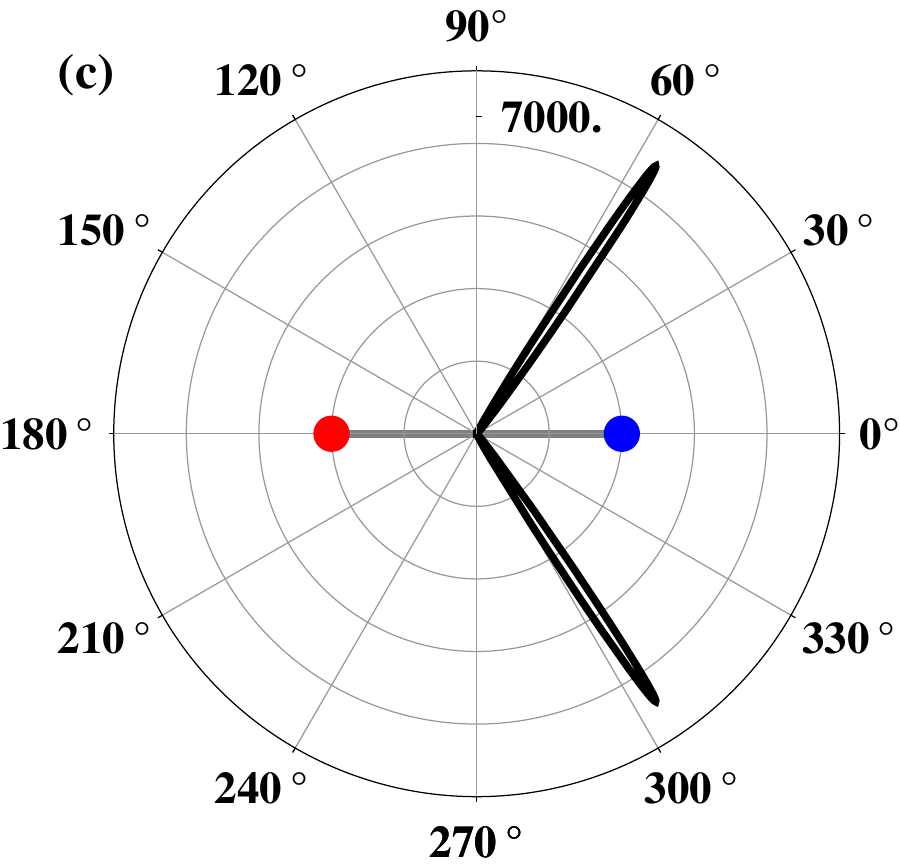}
 \includegraphics[width=.48\columnwidth]{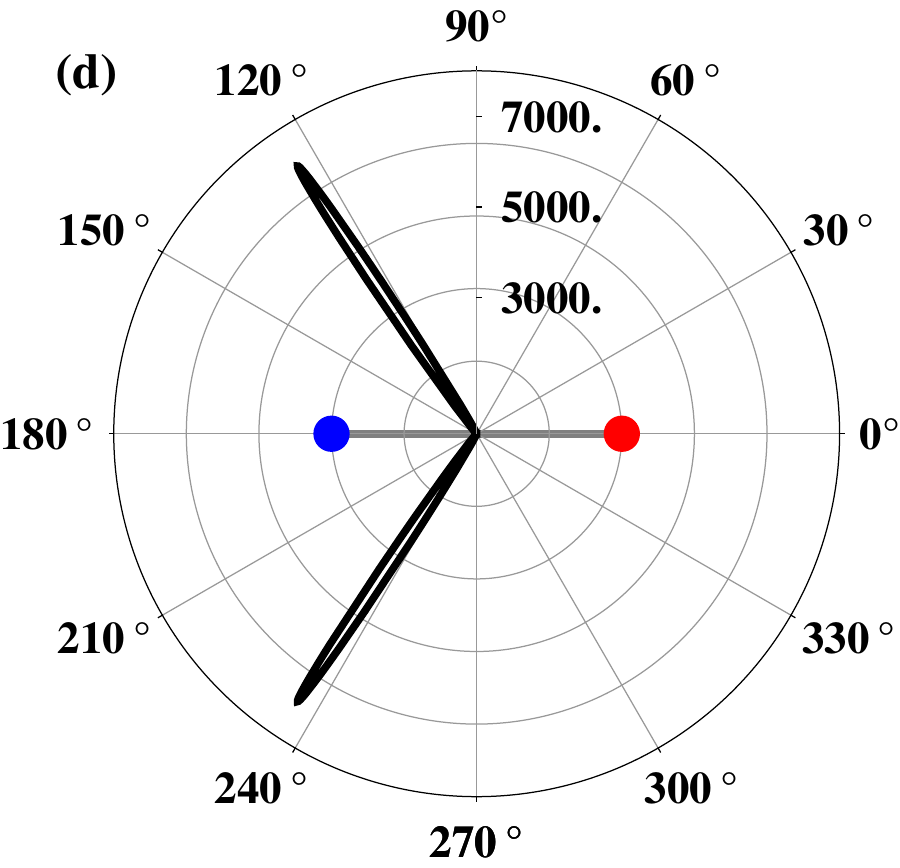}
  \caption{Angular distribution of $g^{(2)}(\vec{R},\vec{R})$ for a chain composed of two atoms illustrated for two different separations between the atoms and two different excitation configurations. In frames (a) and (b), $r_{12}=\lambda/4$, and in frames (c) and (d), $r_{12}=\lambda/2$.
 Frames (a) and (c) show angular distributions corresponding to excitation of left side-atom with a laser field of the Rabi frequency $\Omega=0.02\gamma$. Frames (b) and (d) correspond to excitation of the right-side atom with the same Rabi frequency.}
  \label{fig2}
\end{figure}

Consider now the second-order correlation function $G^{(2)}(\vec{R_{1}},t_{1};\vec{R}_{2},t_{2})$. It is not difficult to see from Eq.~(\ref{q14}) that in the case of two atoms the correlation function can exhibit cosine modulation only if  measured by two distinguishable detectors located at two different geometric points. A simple calculation gives 
\begin{align}
G^{(2)}(\vec{R}_{1},\vec{R}_{2}) &= u(\hat{R}_{1})u(\hat{R}_{2}) \gamma^2\langle S_{1}^{+}S_{2}^{+}S_{1}^{-}S_{2}^{-}\rangle \nonumber\\
&\times\left\{1+\cos\left[k\,\vec{r}_{12}\cdot\left(\hat{R}_{1}-\hat{R}_{2}\right)\right]\right\} .
\end{align}
It is noted that the second-order correlation function manifests an interference pattern dependent on the separation between the two detection positions. The visibility of the interference pattern is independent of the way the atoms are excited. 

However, in the case when the measurement is made with a single detector or two detectors recording in sync, $G^{(2)}(\vec{R}_{1},\vec{R}_{2})=G^{(2)}(\vec{R},\vec{R})$ becomes independent of the direction of detection such that there is no interference pattern in the second-order sense. Thus, simultaneous emission of two photons is spherically symmetric. In other words, photons emitted simultaneously in the same direction do not interfere.

Comparing the properties of $G^{(2)}(\vec{R},\vec{R})$ with those of $G^{(1)}(\vec{R})$ we see that two photons can be detected anywhere despite the fact that one photon can never be detected in certain directions. This fact that one photon can never be detected or can be detected but only with a very small probability in certain directions may result in the {\it superbunching} effect such that the normalized second-order correlation function given by
\begin{align}
g^{(2)}(\vec{R},\vec{R}) &=\frac{G^{(2)}(\vec{R},\vec{R})}{G^{(1)}(\vec{R})G^{(1)}(\vec{R})} ,
\end{align}
could have very large values in directions at which $G^{(1)}(\vec{R})$ is very small.

To illustrate this, we consider the angular distribution of $g^{(2)}(\vec{R},\vec{R})$, which, with the result (\ref{q12}) takes the form
\begin{align}
g^{(2)}(\vec{R},\vec{R}) &= \frac{\eta_{1212}}{\left[1+\upsilon_{12}\cos\left(k\,r_{12}\cos\theta -\psi_{12}\right)\right]^{2}}\, ,\label{q24}
\end{align}
where
\begin{align}
\eta_{1212} = \frac{4\langle S_{1}^{+}S_{2}^{+}S_{1}^{-}S_{2}^{-}\rangle}{\left(\langle S_{1}^{+}S_{1}^{-}\rangle +\langle S_{2}^{+}S_{2}^{-}\rangle\right)^{2}} 
\end{align}
is the second-order coherence between the atoms 1 and 2. It is vivid that in the case of a large degree of the first order coherence between the atoms $(\upsilon_{12}\approx 1)$, the correlation function $g^{(2)}(\vec{R},\vec{R})$ can be very large or even infinite for some directions $\theta$. 
This indicates that in these directions two photons are simultaneously emitted with the absence of single photon emission. Although $g^{(2)}(\vec{R},\vec{R})$ is mostly regarded as a measure of photon-photon correlations, it is evident from Eq.~(\ref{q24}) that $g^{(2)}(\vec{R},\vec{R})$ provides much more information about single photon emissions, determined by $[G^{(1)}(\vec{R})]^{2}$, than about two-photon correlations, determined by $G^{(2)}(\vec{R},\vec{R})$. We will return to this issue later in Sec.~\ref{sec4}.
\begin{figure}[h]
  \includegraphics[width=.8\columnwidth]{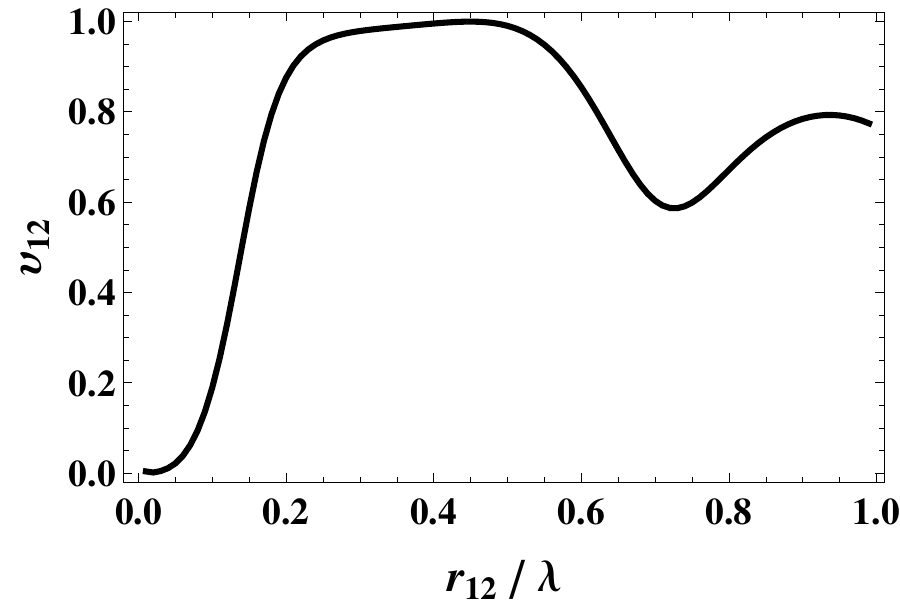}
  \caption{Variation of the degree of the first-order coherence $\upsilon_{12}$ with the scaled inter atomic separation $r_{12}/\lambda$ for the case of the left-side atom driven by a laser field of Rabi frequency $\Omega=0.02\gamma$.}
  \label{fig3}
\end{figure}  

Figure~\ref{fig2} shows the angular distribution pattern of $g^{(2)}(\vec{R},\vec{R})$ for two fixed separations between the atoms $r_{12}=\lambda/4$ and $r_{12}=\lambda/2$, and for two different excitation configurations. The shape of the radiation pattern is very simple. Under excitation of the left-sided atom, the pattern of $g^{(2)}(\vec{R},\vec{R})$ shows two pronounced correlation peaks (superbunching) spatially concentrated in the right half of the pattern, the half of the pattern in which the undriven atom is located. The directions of these peaks point precisely where $G^{(1)}(\vec{R})$ has optimal minima corresponding to an extremely small probability of emission of single photons. The magnitude and sharpness of the peaks depend on the distance between the two atoms. For $r_{12}=\lambda/2$ the peaks are more narrowed, needle shaped, and have magnitudes much larger than ones for $r_{12}=\lambda/4$. When the laser excitation is turned on to the right-sided atom, the correlation pattern flips over the vertical axis or equivalently rotates by $\pi$ radians. Therefore, it is the way the atoms are excited which accounts for the qualitative change of the patterns. In other words, there are preferred directions of no single photon emission imposed by the excitation field. The superbunching effect results from a nonzero phase shift $\psi_{12}$ and thus from the creation of minima of $G^{(1)}(\vec{R},t)$ which is a proof that the single photon emission is significantly suppressed. With the parameter values of Fig.~\ref{fig2}(a), $r_{12}=\lambda/4$ and the weak laser excitation turned on to the left-side atom driven with the Rabi frequency $\Omega=0.02\gamma$, ${\rm Re}[\langle S_{1}^{+}S_{2}^{-}\rangle] \approx -0.000134$ and ${\rm Im}[\langle S_{1}^{+}S_{2}^{-}\rangle]=-0.000297 $, so that $\psi_{12}\approx-0.64\pi$. When the excitation is turned on the right-sided atom, the case illustrated in Fig.~\ref{fig2}(b), ${\rm Im}[\langle S_{1}^{+}S_{2}^{-}\rangle]$ reverses sign and thus $\psi_{12}$ turns out to be $0.64\pi$.

One might argue that a larger number of correlation peaks could be witnessed in the angular distribution of two-photon correlated emission probability pattern when the two atoms are well-separated, that is, when $r_{12}>\lambda/2$. However, for $r_{12}>\lambda/2$, the degree of the first-order coherence $\upsilon_{12}$ is considerably reduced so that there is no significant reduction of $G^{(1)}(\vec{R})$ present and no subsequent superbunching is possible. Figure~\ref{fig3} illustrates the variation of $\upsilon_{12}$ with $r_{12}/\lambda$. It is apparent that $\upsilon_{12}\approx 1$ for atomic separations $1/4< r_{12}/\lambda<1/2$. Thus, in the case of a bi-atomic chain with the laser excitation driving only one of the atoms in the chain, an almost perfect coherence between the atomic dipole moments is possible to be achieved for inter  atomic spacings $r_{12}\leq\lambda/2$.

\subsection{Atomic chain composed of three atoms}

When the chain is composed of three atoms, both the first and second-order correlation functions depend on the direction of detection such that there is an interference pattern not only in the first-order but also in the second-order sense.

In the case of three atoms the angular distribution of the first-order correlation function is of the form
\begin{align}
\frac{G^{(1)}(\vec{R})}{u(\hat{R})} &= \left(I_{1}+I_{2}\right)\left[\frac{1}{2}+\upsilon_{12}\cos\left(k\,r_{12}\cos\theta -\psi_{12}\right)\right] \nonumber\\
&+\left(I_{2}+I_{3}\right)\left[\frac{1}{2}+\upsilon_{23}\cos\left(k\,r_{23}\cos\theta -\psi_{23}\right)\right] \nonumber\\
&+ \left(I_{3}+I_{1}\right)\left[\frac{1}{2}+\upsilon_{31}\cos\left(k\,r_{31}\cos\theta -\psi_{31}\right)\right] ,\label{q26}
\end{align}
where $I_{i}=\langle S_{i}^{+}S_{i}^{-}\rangle$, $\upsilon_{ij}$ is the first-order coherence between atoms $i$ and $j$, and $\psi_{ij}={\rm arg}(\langle S_{i}^{+}S_{j}^{-}\rangle)$. The correlation function is composed of three terms resulting from the three possible pairs of atoms forming the three-atom chain. 

The angular distribution of the second-order correlation function has the form
\begin{align}
\frac{G^{(2)}(\vec{R},\vec{R})}{4u^{2}(\hat{R})} &= G_{1212}+G_{2323}+G_{3131} \nonumber\\
& +2|G_{1312}|\cos\left(k\,r_{12}\cos\theta -\phi_{12}\right) \nonumber\\
& +2|G_{2313}|\cos\left(k\,r_{23}\cos\theta -\phi_{23}\right) \nonumber\\
& +2|G_{3221}|\cos\left(k\,r_{31}\cos\theta -\phi_{31}\right) ,\label{q27}
\end{align}
where $G_{ijkl}=|\langle S_{i}^{+}S_{j}^{+}S_{k}^{-}S_{l}^{-}\rangle|$ and $\phi_{il}={\rm arg}(G_{ijkl})$. In writing the above expression we have used the fact that $G_{1213}=G^{\ast}_{1312}$, $G_{2312}=G^{\ast}_{1223}$, and $G_{1323}=G^{\ast}_{2313}$.
\begin{figure}[ht]
\includegraphics[width=.45\columnwidth]{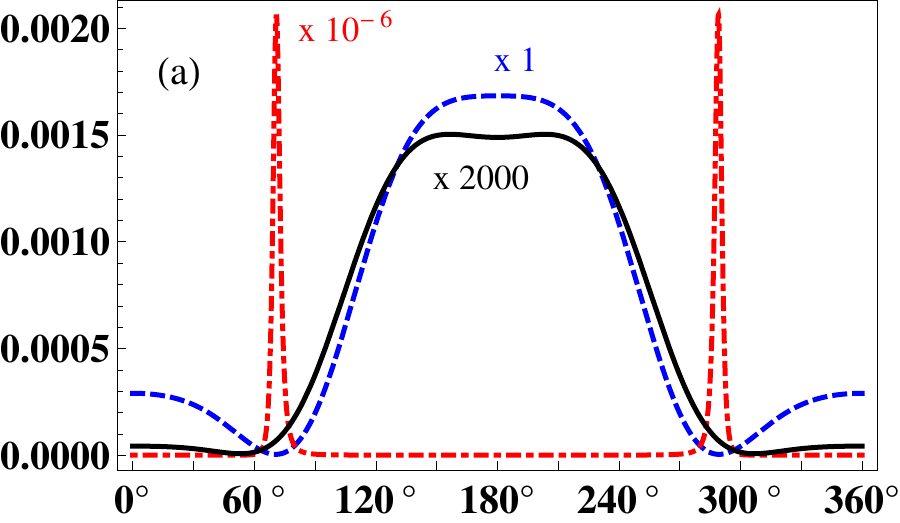}
  \includegraphics[width=.45\columnwidth]{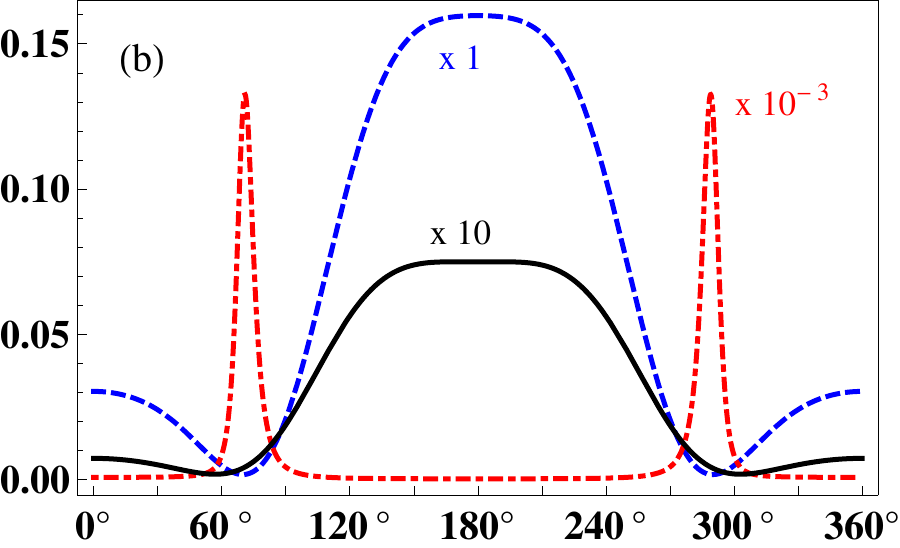}
   \includegraphics[width=.45\columnwidth]{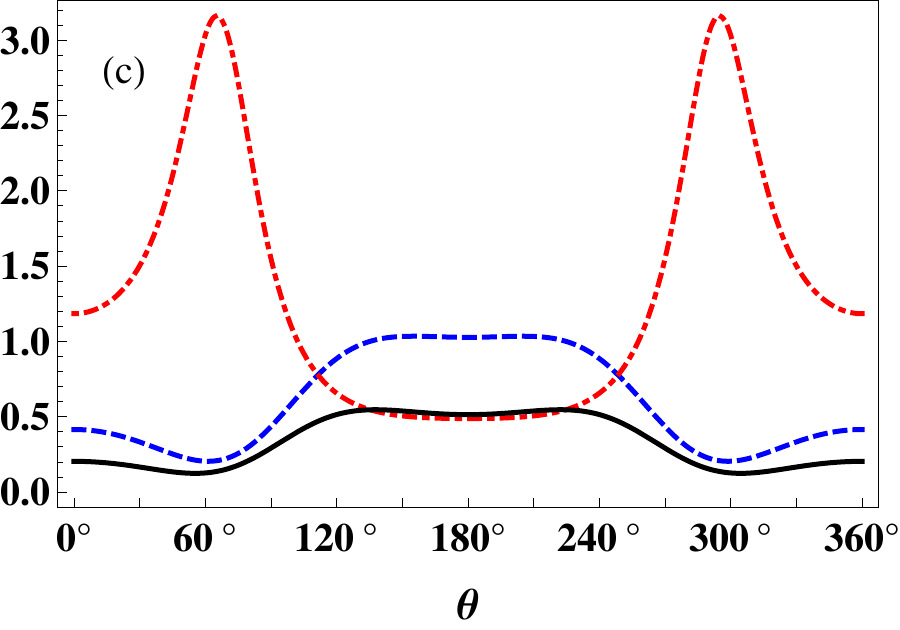}
    \includegraphics[width=.45\columnwidth]{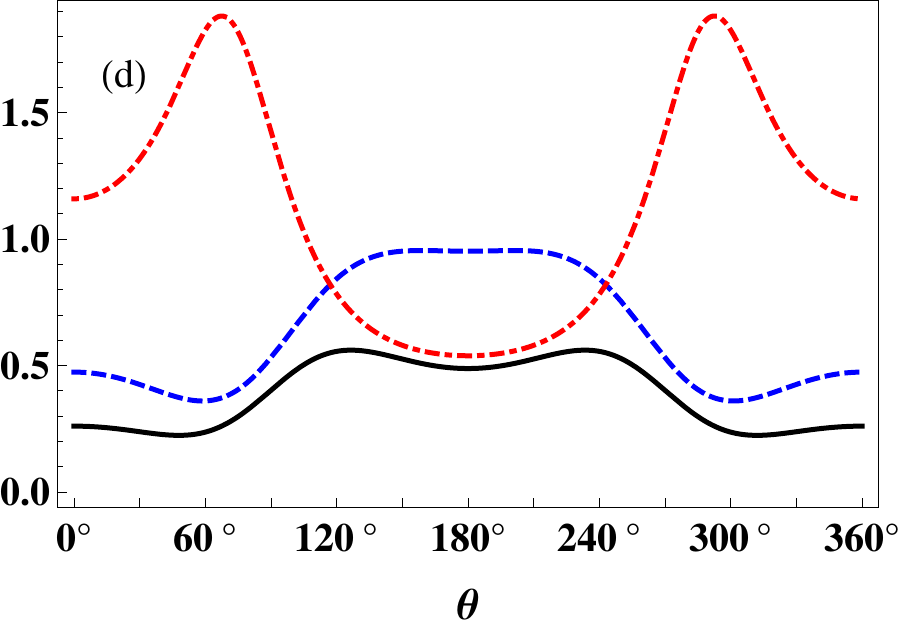}
  \caption{Angular distribution of $G^{(2)}(\vec{R},\vec{R})/[u(\hat{R})]^2$ (solid black line) and $G^{(1)}(\vec{R})/u(\hat{R})$ (dashed blue line) for $r_{12}=r_{23}=\lambda/4$ plotted against detection angle $\theta$. Different Rabi frequencies of the laser field drive the left-sided atom, (a) $\Omega=0.02\gamma$, (b) $\Omega=0.2\gamma$, (c) $\Omega=\gamma$, and (d) $\Omega=10\gamma$. Also shown is the normalized second-order correlation function $g^{(2)}(\vec{R},\vec{R})$ (dashed-dotted red line). The curves in (a) and (b) have been scaled with constant factors. }
  \label{fig4}
\end{figure}

If we introduce the abbreviations
\begin{align}
G_{1} &\equiv G_{1212} ,\quad G_{2}\equiv G_{2323} ,\quad G_{3}\equiv G_{3131} ,\nonumber\\
\sigma_{12} &=\frac{2|G_{1312}|}{G_{1}+G_{2}} ,\ \sigma_{23} =\frac{2|G_{2313}|}{G_{2}+G_{3}} ,\ \sigma_{31} =\frac{2|G_{3221}|}{G_{3}+G_{1}} ,\label{q28}
\end{align}
then we can write the second-order correlation function divided by the prefactor, Eq.~(\ref{q27}), as
\begin{align}
\frac{G^{(2)}(\vec{R},\vec{R})}{4u^{2}(\hat{R})} &= \left(G_{1}+G_{2}\right)\left[\frac{1}{2}+\sigma_{12}\cos\left(k\,r_{12}\cos\theta -\phi_{12}\right)\right] \nonumber\\
&+\left(G_{2}+G_{3}\right)\left[\frac{1}{2}+\sigma_{23}\cos\left(k\,r_{23}\cos\theta -\phi_{23}\right)\right] \nonumber\\
&+ \left(G_{3}+G_{1}\right)\left[\frac{1}{2}+\sigma_{31}\cos\left(k\,r_{31}\cos\theta -\phi_{31}\right)\right] .\label{q29}
\end{align}
Note that $\sigma_{ij}$ do not represent the correlation coefficients in the same sense as the first-order coherence $\upsilon_{ij}$. They represent some kind of correlations but do not necessarily obey $\sigma_{ij}\leq 1$.
\begin{figure}[t]
 \includegraphics[width=.48\columnwidth]{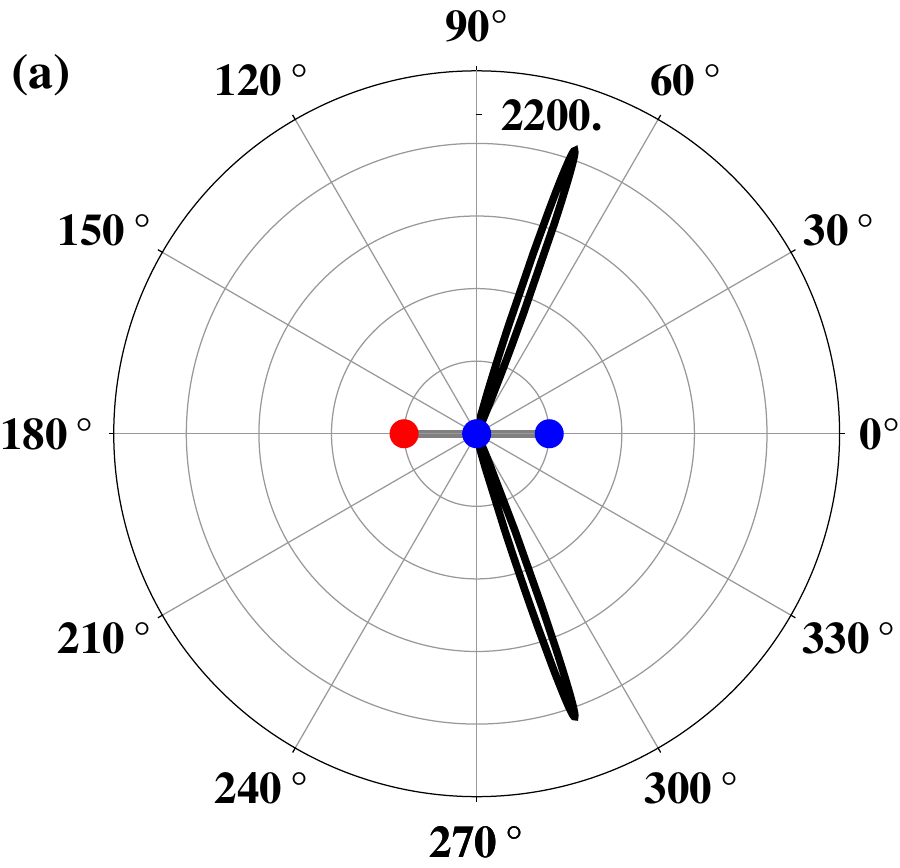}
   \hspace{0.2 cm}
    \includegraphics[width=.48\columnwidth]{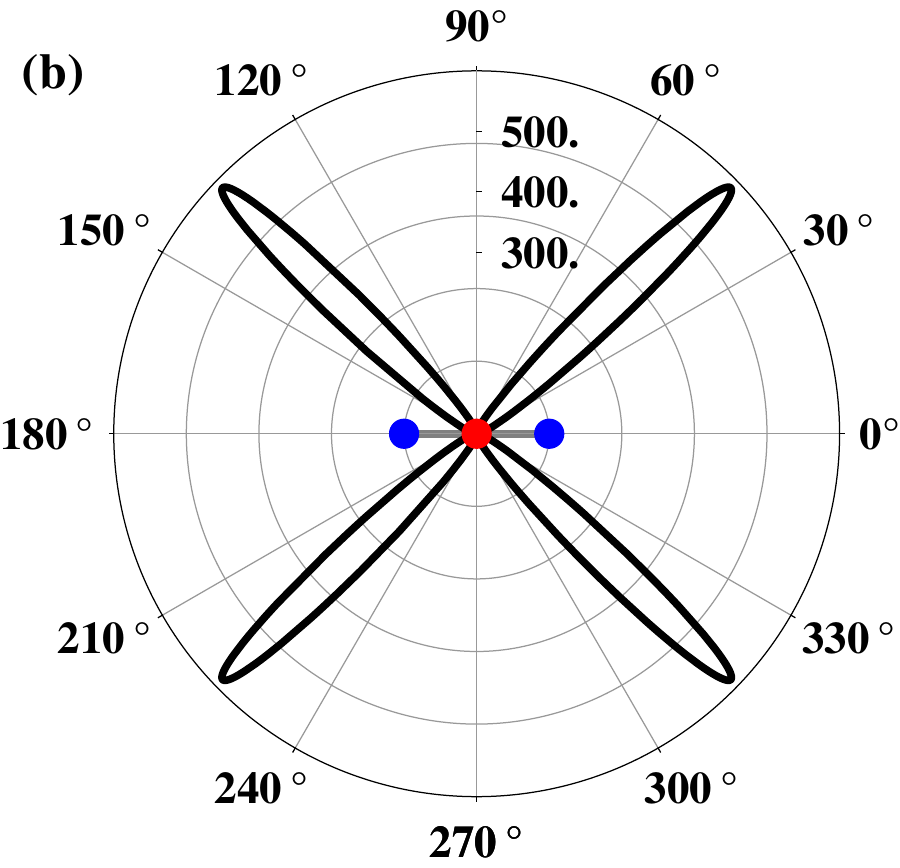}
     \includegraphics[width=.48\columnwidth]{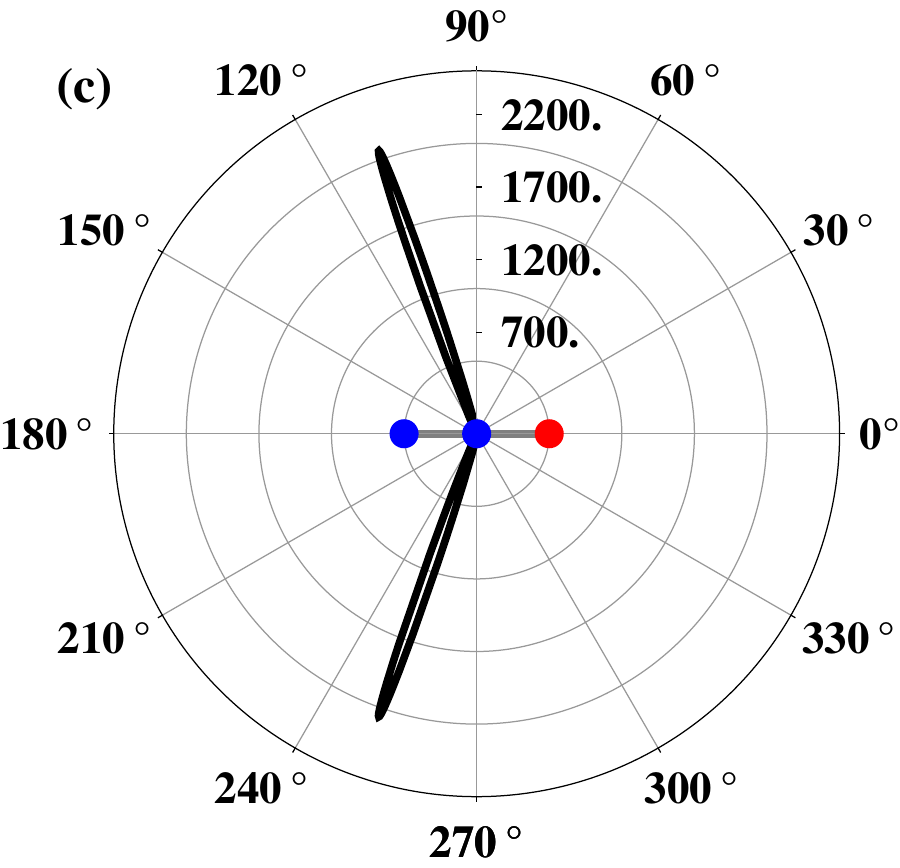}
  \caption{Angular distribution of $g^{(2)}(\vec{R},\vec{R})$ for a chain composed of three atoms with $r_{12}=r_{23}=\lambda/4$ and for different excitation configurations, (a) left-sided atom, (b) middle atom, and (c) right-sided atom driven by a laser field of the Rabi frequency $\Omega=0.02\gamma$.}
  \label{fig5}
\end{figure}

It is interesting that $G^{(1)}(\vec{R})$, Eq.~(\ref{q26}),  and $G^{(2)}(\vec{R},\vec{R})$, Eq.~(\ref{q29}), are so analogous in appearance. This fact hints that the angular distribution of $G^{(2)}(\vec{R},\vec{R})$ is expected to be similar in form to that of $G^{(1)}(\vec{R})$, except that the magnitudes and the directions of maxima and minima might be different. 

The angular distributions of the correlation functions are shown in Fig.~\ref{fig4}, where we plot $G^{(2)}(\vec{R},\vec{R})/[u(\hat{R})]^2$ and $G^{(1)}(\vec{R})/u(\hat{R})$ for several different values of the Rabi frequency of the laser field driving the left-sided atom of the chain. It is clearly seen that the angular distributions of the correlations are similar in form. Both the first and second-order correlations are 
maximal in the direction $\theta=\pi$, the backward direction relative to the direction of the chain. However, in these directions $G^{(1)}(\vec{R})$ is larger than $G^{(2)}(\vec{R},\vec{R})$ indicating antibunching of the emitted photons. Correlated pairs of photons with $G^{(2)}(\vec{R},\vec{R})$ comparable and even larger that $G^{(1)}(\vec{R})$ are emitted in directions located on that side of the pattern where the undriven atoms are located.

In Fig.~\ref{fig5} we show polar diagrams of the $g^{(2)}(\vec{R},\vec{R})$ function for a chain composed of three atoms with $r_{12}=r_{23}=\lambda/4$ and for different laser excitation configurations. It is seen that the direction and the number of correlation (superbunching) peaks depend on the driving field configuration. When one of the side atoms is driven,  Fig.~\ref{fig5}(a) and (c), the pattern exhibits two pronounced peaks spatially concentrated in that half of the detection plane where the undriven atoms are located.  
Comparing the results for the chain composed of three atoms with those for two atoms, Fig.~\ref{fig2}, we see that the magnitude of the correlation peaks increases and the peaks become  narrower when the number of atoms in the chain is increased. Thus, a longer chain not only generates a stronger superbunched light but also leads to a better location of the strongly correlated pairs of photons. 
When the middle atom is driven, the pattern is composed of four pronounced peaks.
It is easy to understand why four instead of two peaks appear in the pattern. When the middle atom is driven the total system is equivalent to the case of two atomic sub-chains each composed of two atoms. In the sub-chain composed of atoms $1$ and $2$, the right-sided atoms is driven whereas in the sub-chain composed of atoms $2$ and $3$ the left-sided atom is driven. Each sub-chain produces two correlation peaks located in opposite half of the pattern.

\begin{figure}[t]
  \centerline{\includegraphics[width=230pt]{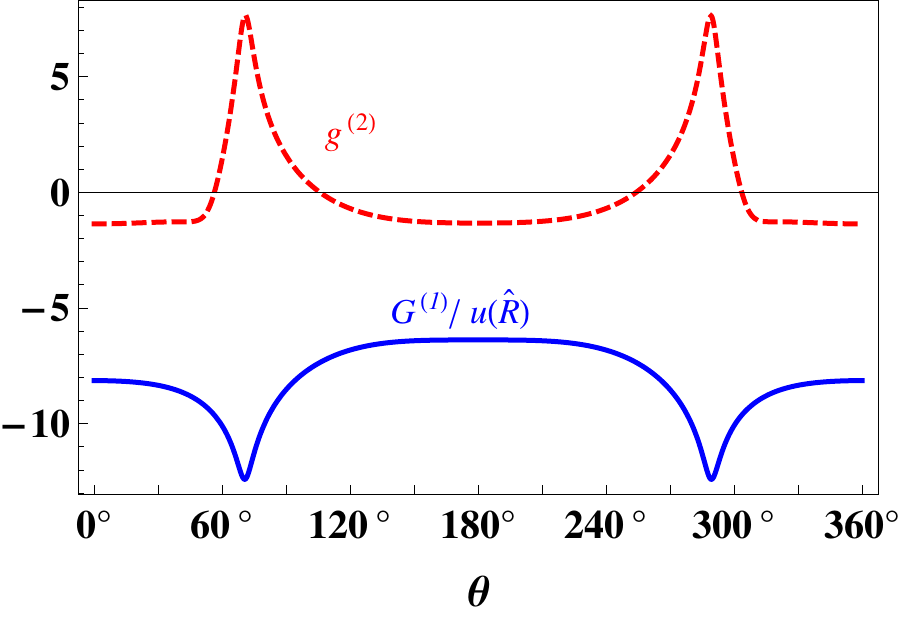}}
  \caption{Comparison of the angular distributions of the correlation function $G^{(1)}(\vec{R})/u(\hat{R})$ (solid blue line) with that of $g^{(2)}(\vec{R},\vec{R})$ (dashed red line) plotted using logarithmic scale against the detection angle $\theta$. The  parameters are same as in Fig.~\ref{fig5}(a).}
  \label{fig6}
\end{figure}

It follows from Fig.~\ref{fig4}(a), which corresponds to the same situation as shown in Fig.~\ref{fig5}(a), that at the directions of the superbunched peaks $G^{(1)}(\vec{R})$ reaches its optimal minimum.  Figure~\ref{fig6} shows the angular distributions of the three terms composing of $G^{(1)}(\vec{R})$ plotted for the same parameters as in Fig.~\ref{fig5}(a). It is clearly seen that superbunching occurs in the directions in which the probability of the emission of single photons is nearly zero.
\subsection{Atomic chains composed of $N>3$ atoms}

\begin{figure}[t]
\includegraphics[width=0.48\columnwidth]{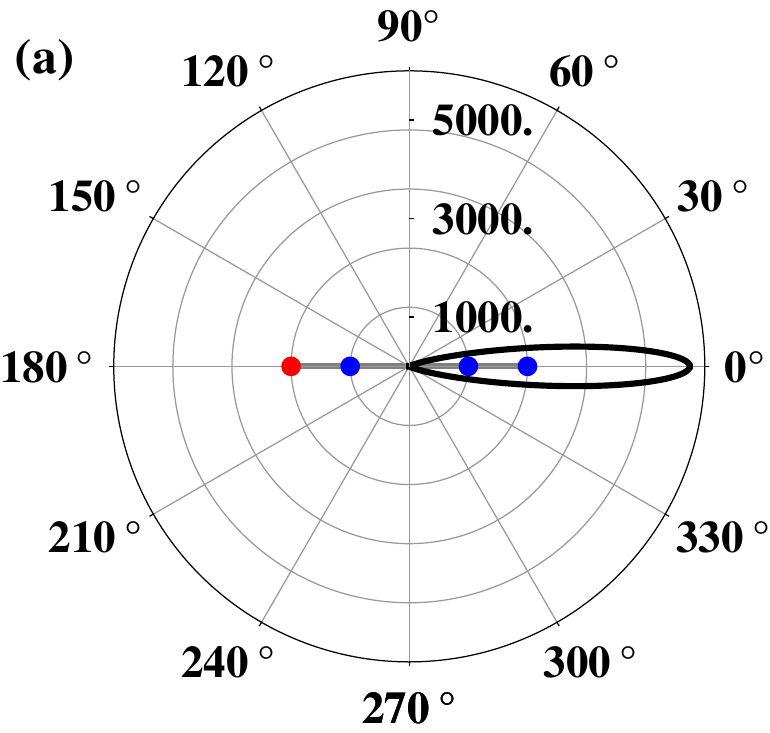}
\includegraphics[width=0.47\columnwidth]{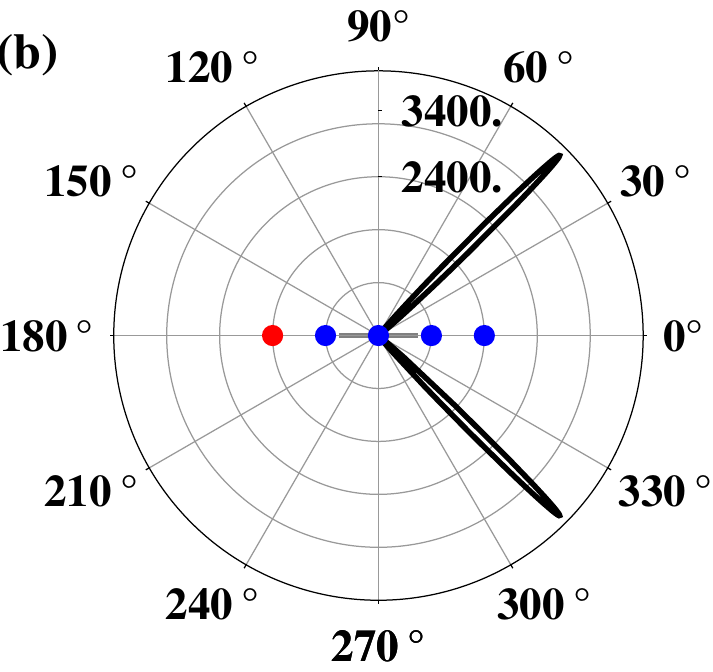}
  \caption{Angular distribution of $g^{(2)}(\vec{R},\vec{R})$ for a chain composed of (a) $N=4$, (b) $N=5$ equidistant atoms with $r_{ij}=\lambda/4$. Left-sided atom driven by a laser field of the Rabi frequency $\Omega=0.02\gamma$..}
  \label{figbig}
\end{figure}

Finally, our analysis is supplemented by the inclusion of the cases where the atomic chains contain more than three atoms. In Fig.~\ref{figbig}, we show polar diagrams of the normalized intensity correlation function with $N=4$ and $N=5$ atoms for $r_{ij}=\lambda/4$, respectively. One can immediately notice the raise in the magnitude of the peaks indicating super bunching of the scattered photons compared to the two and three atoms cases. However, it becomes important here to notice that the increased magnitude has to be compared separately for even and odd values of $N$. That is to say, Fig.~\ref{figbig}(a) should be compared with Fig.~\ref{fig2}(a) and Fig.~\ref{figbig}(b) should be contrasted against Fig.~\ref{fig5}(a). Moreover, another prompt observation by the comparison of these figures is that the directions of emission of strongly correlated pairs of photons turn toward the inter atomic axis. This turning of super bunching peaks is more drastic when the total number of atoms in the chain is even than when it is odd, i.e., in Fig.~\ref{figbig}(a), for four atoms comprising the atomic chain, the two peaks merge and appear as one single peak, and lie already parallel to $\theta=0^{\circ}$ or to the line of atoms.

Hence, one can assert that the effect of increasing the number of atoms in the chain is to reinforce the superbunching of the emitted photon pairs and the corresponding directions tend to bend toward the atomic chain. This happens more rapidly if $N$ is an even number.

\section{Comments on the meaning of superbunching}\label{sec4}

The term superbunching is used in general for $g^{(2)}(\vec{R},\vec{R})\gg 1$ and is interpreted as a signature of strong photon-photon correlations. 
The considerations of Sec.~\ref{sec3} show that the question of whether superbunching means strong photon-photon correlations may be irrelevant to the problem of obtaining large values of $g^{(2)}(\vec{R},\vec{R})$ at directions where single photons are not emitted. Let us illustrate this point more clearly. 
\begin{figure}[t]
  \centerline{\includegraphics[width=250pt]{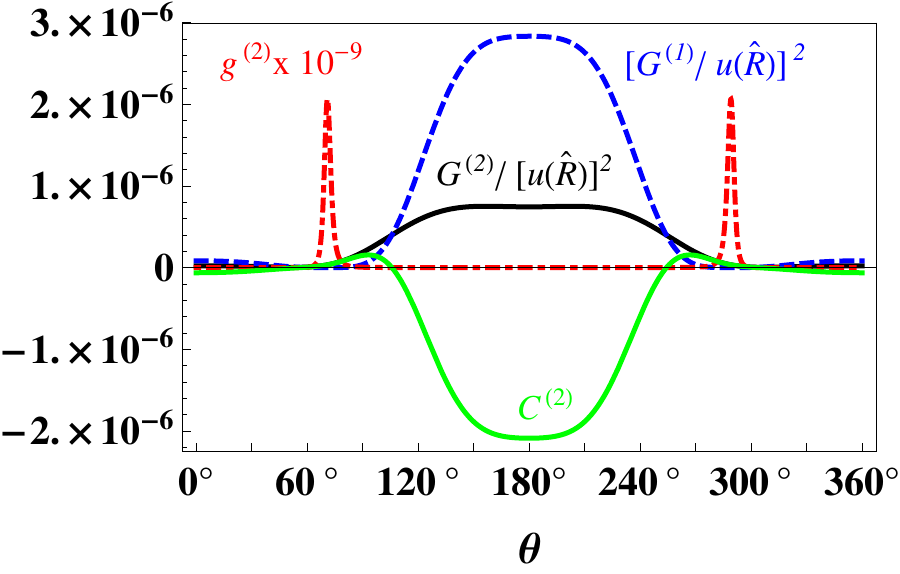}}
	  \caption{Angular distributions of $G^{(2)}(\vec{R},\vec{R})/(u(\hat{R}))^2$ (solid black line), $[G^{(1)}(\hat{R})/u(\hat{R})]^{2}$ (dashed blue line) and $g^{(2)}(\vec{R},\vec{R})*10^{-9}$ (dashed-dotted red line)  for a chain composed of three atoms separated by $r_{12}=r_{23}=\lambda/4$  plotted against the detection angle $\theta$. The left-sided atom is driven by a laser field of the Rabi frequency $\Omega=0.02\gamma$. Also shown is a function $C^{(2)}(\vec{R},\vec{R}) = G^{(2)}(\vec{R},\vec{R})/(u(\hat{R}))^2 -[G^{(1)}(\vec{R})/u(\hat{R})]^{2}$ (solid green line).}
  \label{fig7}
\end{figure}

In Fig.~\ref{fig7} we show the angular distribution of the correlation functions $[G^{(1)}(\vec{R})]^{2}$, $G^{(2)}(\vec{R},\vec{R})$, and $g^{(2)}(\vec{R},\vec{R})$ for the case of three atoms separated by $r_{12}=r_{23}=\lambda/4$. It is seen that at the angles where superbunching occurs $G^{(1)}(\vec{R})$ reaches its optimal minimum and also $G^{(2)}(\vec{R},\vec{R})$ is significantly reduced, close to its optimal minimum. Thus, at the angles where the maximum superbunching occurs, i.e., $71^{\circ}$ and $289^{\circ}$ in this case, $G^{(2)}(\vec{R},\vec{R})$ is pretty small.  $G^{(2)}(\vec{R},\vec{R})/(u(\hat{R}))^2\sim 3.5*10^{-8} $. Simultaneously, $[G^{(1)}(\vec{R})/u(\hat{R})]^{2}\sim 1.7*10^{-11}$, so that the ratio $G^{(2)}(\vec{R},\vec{R})/[G^{(1)}(\vec{R})]^{2}$ is very large. This shows that $g^{(2)}(\vec{R},\vec{R})$ varies much more vigoroulsy with $[G^{(1)}(\vec{R})]^{2}$ than with  $G^{(2)}(\vec{R},\vec{R})$. Thus, $g^{(2)}(\vec{R},\vec{R})$ could be regarded as a better measure of $[G^{(1)}(\vec{R})]^{2}$ rather than $G^{(2)}(\vec{R},\vec{R})$. In other words, $g^{(2)}(\vec{R},\vec{R})$ provides much more information about $[G^{(1)}(\vec{R})]^{2}$ than about $G^{(2)}(\vec{R},\vec{R})$.

One can also see from Fig.~\ref{fig7} that at the angles $\theta$ at which $g^{(2)}(\vec{R},\vec{R})$ is maximal, the difference between $G^{(2)}(\vec{R},\vec{R})$ and $[G^{(1)}(\vec{R})]^{2}$ is also maximal. Therefore, instead of $g^{(2)}(\vec{R},\vec{R})$, we may consider a correlation measure defined in~\cite{remp}
\begin{align}
C^{(2)}(\vec{R},\vec{R})(u(\hat{R}))^2 &= G^{(2)}(\vec{R},\vec{R}) -\left[G^{(1)}(\vec{R})\right]^{2} \nonumber\\
&=\left\{g^{(2)}(\vec{R},\vec{R}) -1\right\}\left[G^{(1)}(\vec{R})\right]^{2} .
\end{align}
The correlation function $C^{(2)}(\vec{R},\vec{R})(u(\hat{R}))^2 $ is less sensitive to single-photon emissions than $g^{(2)}(\vec{R},\vec{R})$ and provides a clearer measure of $G^{(2)}(\vec{R},\vec{R})$, the probability of the simultaneous emission of two photons.
Positive values of $C^{(2)}(\vec{R},\vec{R})(u(\hat{R}))^2 $ indicate that simultaneous emission of two photons dominates over the single photon emissions, $C^{(2)}(\vec{R},\vec{R})(u(\hat{R}))^2 =0$ corresponds to a coherent emission and $C^{(2)}(\vec{R},\vec{R})(u(\hat{R}))^2 <0$ indicates emission of single photons, with the minimum negative value $C^{(2)}(\vec{R},\vec{R})(u(\hat{R}))^2 = -G^{(1)}(\vec{R})G^{(1)}(\vec{R})$ corresponding to the emission of a single photon. It is evident from Fig.~\ref{fig7} that $C^{(2)}(\vec{R},\vec{R})(u(\hat{R}))^2 $ is less responsive to $[G^{(1)}(\vec{R})]^{2}$ than $g^{(2)}(\vec{R},\vec{R})$ and provides information about values of $G^{(2)}(\vec{R},\vec{R})$ even if $G^{(1)}(\vec{R})=0$.

Following the above analysis, we may conclude that superbunching as determined by $g^{(2)}(\vec{R},\vec{R})\gg 1$, implies that the probability of the emission of two single photons is much smaller rather than the preconceived notion of strong photon-photon correlations or alternatively, the probability of the synchronized emission of two photons at the same time. 

\section{Summary}
\label{sec5}

In this paper we have studied the correlation characteristics of the fluorescence field emitted from a linear chain of identical two-level atoms. It has been assumed that only one of the atoms of the chain is selectively driven by a coherent laser field. The atoms interact with each other through the dipole-dipole interaction and the collective spontaneous emission resulting from the coupling of the atoms to a common vacuum field. In such a system the interference pattern of the radiation field and the correlations between the atoms depend not only on the physical  geometry of the system (distance between the atoms) but also on the arrangement of driven atom. We have found that the effect of selective driving of only a single atom results in a shift of the phase difference between neighboring atoms. The shift leads to a destructive interference of the emitted radiation that significantly reduces the probability of emission of single photons. The immediate effect of the reduced single photon emission is to produce pronounced peaks in the angular distribution of the normalized second-order correlation function. The maximum value of these peaks can be made huge, values of the order of hundreds or thousands, which is termed as superbunching. When one of the side-most atom is driven by the laser and the separation between atoms is kept less than or equal to half of the resonant atomic transition wavelength, the normalized second-order correlation function exhibits single or two superbunched peaks. When the driving field is turned on to the other side-most atom, the directions of the superbunched peaks flips by $\pi$ radians. Switching the driving field on to the middle atom of the chain results in two or four superbunched peaks in the angular distribution of the normalized second-order correlation function. The effect of increasing the number of emitters in the chain is to produce more prominent superbunched peaks which tend to turn toward the atomic line. The meaning of superbunching has also been discussed and we have argued that the normalized second-order correlation function, which is regarded as a measure of photon-photon correlations, provides much more information about single photon emission than about simultaneous two-photon emission.


\begin{thebibliography}{21}

\bibitem{coldgas} I. Bloch, J. Dalibard, and W. Zwerger, Rev. Mod. Phys. {\bf 80}, 885 (2008).

\bibitem{Ni} K.-K. Ni {\it et al.}, Science {\bf 322}, 231 (2008).

\bibitem{ryd1} M. Saffman, T. G. Walker and K. Molmer, Rev. Mod. Phys. {\bf 82}, 2313 (2010).

\bibitem{ryd2} D. Comparat and P. Pillet,  J. Opt. Soc. Am. B {\bf 27}, A208 (2010).

\bibitem{nm07} K. P. Nayak, P. N. Melentiev, M. Morinaga, F. Le Kien, V. I. Balykin, and K. Hakuta, Opt. Express {\bf 15}, 5431 (2007).

\bibitem{vr10} E. Vetsch, D. Reitz, G. Sagu\'{e}, R. Schmidt, S. T. Dawkins, and A. Rauschenbeutel, Phys. Rev. Lett. {\bf 104}, 203603 (2010). 

\bibitem{coldgasImaging} C. Weitenberg {\it et al.}, Phys. Rev. Lett. {\bf 106}, 215301 (2011).

\bibitem{agarwal} G. S. Agarwal, Springer Tracts in Modern Physics: Quantum Optics (Springer-Verlag, Berlin, 1974).

\bibitem{ficekbook}
Z. Ficek and S. Swain, {\it Quantum Interference and Coherence: Theory and Experiments}, Vol. 100, (Springer, New York, 2005).

\bibitem{scullybook}M. O. Scully and M. S. Zubairy, {\it Quantum Optics}, (Cambridge University Press, Cambridge, 1997). 

\bibitem{d54} R. H. Dicke, Phys. Rev.   {\bf 93}, 99 (1954).

\bibitem{Lehmberg} R. H. Lehmberg, Phys. Rev. A {\bf 2}, 889 (1970).

\bibitem{bb84} H. Blank, M. Blank, K. Blum, and A. Faridani, Phys. Lett. {\bf 105A}, 39 (1984).

\bibitem{f86} H. S. Freedhoff, J. Chem. Phys. {\bf 85}, 6110 (1986).

\bibitem{corr1} Z. Ficek, R. Tana\'s, and S. Kielich, Physica {\bf 146A}, 452 (1987).

\bibitem{corr2}S. Das, G. S. Agarwal, and M. O. Scully, Phys. Rev. Lett. {\bf 101}, 153601 (2008).

\bibitem{corr3}J. Eschner {\it et al.}, Nature {\bf 413},  495 (2001); S.  Rist, J. Eschner, M. Hennrich, and G. Morigi, Phys. Rev. A {\bf 78}, 013808 (2008).

\bibitem{corr4} C. Hettich {\it et al.}, Science {\bf 298}, 385 (2002).

\bibitem{distance}J.-T. Chang, J. Evers, M. O. Scully and M. S. Zubairy,
Phys. Rev. A {\bf 73}, 031803(R) (2006); J.-T. Chang, J. Evers and M. S. Zubairy,
Phys. Rev. A {\bf 74}, 043820 (2006); Q. Gulfam and J. Evers, J. Phys. B At. Mol. Opt. Phys. {\bf 43}, 045501 (2010).

\bibitem{Carmichael} J. P. Clemens, L. Horvath, B. C. Sanders, and H. J. Carmichael, Phys. Rev. A {\bf 68}, 023809 (2003).

\bibitem{mf07} C. J. Mewton and Z. Ficek, J. Phys. B: At. Mol. Opt. Phys. {\bf B40}, S181 (2007).

\bibitem{Cirac}D. Porras and J. I. Cirac, Phys. Rev. A {\bf 78}, 053816 (2008).

\bibitem{lz14} Z. Liao and M. S. Zubairy, Phys. Rev. A {\bf 90}, 053805 (2014).

\bibitem{mirror} Q. Gulfam and Z. Ficek, Phys. Rev. A {\bf 94}, 053831 (2016).

\bibitem{Richter} T. Richter, J. Phys. B: At. Mol. Phys. {\bf 15}, 1293 (1982).

\bibitem{richter2} T. Richter, Annalen der Physik {\bf 495},  234 (1983).

\bibitem{antibunching1}A. Kuhn, M. Hennrich, and G. Rempe, Phys. Rev. Lett. {\bf 89}, 067901 (2002).

\bibitem{antibunching2}J. McKeever {\it et al.}, Science {\bf 303}, 1992 (2004).

\bibitem{antibunching3}M. Keller {\it et al.}, Nature {\bf 431},  1075 (2004).

\bibitem{antibunching4}K. M. Birnbaum {\it et al.}, Nature {\bf 436}, 87 (2005).

\bibitem{antibunching5}M. Hijlkema {\it et al.}, Nat. Phys. {\bf 3}, 253 (2007).

\bibitem{antibunching6}B. Dayan {\it et al.}, Science {\bf 319}, 1062 (2008).

\bibitem{Wiegand} M. Wiegand, Opt. Commun. {\bf 36} 297 (1981).

\bibitem{superbunching1}H. J. Carmichael, P. Kochan, and B. C. Sanders, Phys. Rev. Lett. {\bf 77}, 631 (1996).

\bibitem{characterization1}I. Schuster, A. Kubanek, A. Fuhrmanek, T. Puppe, P. W. H. Pinkse, K. Murr and G. Rempe, Nat. Phys. {\bf 4}, 382 (2008).

\bibitem{superbunching2}L. Schneebeli, M. Kira, and S. W. Koch, Phys. Rev. Lett. {\bf 101}, 097401 (2008).

\bibitem{AIP-paper} Q. Gulfam and Z. Ficek, AIP conference proceedings, {\bf 1976}, 020003 (2018).

\bibitem{Wstate} R. Weigner, J. von Zanthier, and G. S. Agarwal, Phys. Rev. A {\bf 84}, 023805 (2011).

\bibitem{Bin}B. Bai, J. Liu, Y. Zhou, H. Zheng, H. Chen, S. Zhang, Y. He, F. Li, and Z. Xu, J. Opt. Soc. Am. B {\bf 34}, 2081 (2017).

\bibitem{Leuchs}T. Sh. Iskhakov, A. M. Pérez, K. Y. Spasibko, M. Chekhova, and G. Leuchs, Opt. Lett. {\bf 37}, 1919 (2012).  

\bibitem{walls}H. J. Carmichael and D. F. Walls, J. Phys. B {\bf 9}, 1199 (1976).

\bibitem{kl76} H. J. Kimble and L. Mandel, Phys. Rev. A {\bf 13}, 2123 (1976).

\bibitem{kl77}H. J. Kimble, M. Dagenais, and L. Mandel, Phys. Rev. Lett. {\bf 39}, 691 (1977).

\bibitem{glauber} R. J. Glauber, Phys. Rev. {\bf 130}, 2529 (1963).

\bibitem{fs} M. Kiffner, M. Macovei, J. Evers, and C. H. Keitel, {\it Vacuum-induced processes in multi-level atoms}, in Progress in Optics Vol. 55, Ed. E. Wolf, Elsevier, Amsterdam, (2010).

\bibitem{remp} A. Kubanek, A. Ourjoumtsev, I. Schuster, M. Koch, P. W. H. Pinkse, K. Murr and G. Rempe, Phys. Rev. Lett. {\bf 101}, 203602 (2008).

\end{thebibliography}
\end{document}